\documentstyle[11pt,paspconf,epsf]{article}

\begin{document}
\title{Main Sequence Fitting and the Hipparcos Open Cluster Distance Scale}
\author{M.\ H. Pinsonneault, D.\ M.\ Terndrup, and Y.\ Yuan}
\affil{Ohio State University, Dept. of Astronomy, 140 W. 18th Ave.,
Columbus, OH 43210}

\begin{abstract}
The difference between the Hipparcos open cluster distance scale and
that obtained from main sequence fitting is examined.  The two color
main sequence fitting technique of Pinsonneault et al.\ (1998) is
extended to NGC 2516, NGC 6475, and NGC 6633.  The error sources for
main sequence fitting are examined, and possible evidence for
systematic errors in the Hipparcos parallaxes is discussed.
\end{abstract}


\section{Introduction}

One of the most difficult and important issues in astronomy is the
determination of distance scales.  Until recently the open cluster
distance scale was not generally regarded as controversial; however,
this changed when the Hipparcos mission permitted the measurement of
trigonometric parallaxes for a number of stars in open clusters more
distant than the Hyades.  The distances obtained from Hipparcos
parallaxes are in significant disagreement with previous distance
measurements obtained from main sequence fitting techniques in some of
the best-studied systems in the sky (for example, the Pleiades and
Praesepe clusters).  This is not a problem with an easy answer; the
technique of main sequence fitting is based upon a firm foundation in
the theory of stellar structure and evolution and is supported by a
wealth of empirical tests and constraints.  On the other hand, the
Hipparcos mission has provided an unprecedented database of fundamental
distance measurements and there is a large body of solid astrometric
work behind the Hipparcos measurements.

In this talk I will begin by summarizing the major issues involved, and
follow with a brief description of main sequence fitting and the
Hipparcos open cluster distance scales.  I will then discuss the
possible error sources in the main sequence fitting techniques and the
possibility of systematic errors in the Hipparcos open cluster
distances.  Finally, I will close with an illustration of some of the
other possible applications of main sequence fitting: namely, an age
diagnostic for young clusters and the ability to assign more precise
relative distances and metal abundances to less well-studied open
clusters.

\subsection{Why Does it Matter?}

The differences between the Hipparcos and main sequence fitting
distance scales might seem to be an arcane issue; however, it
indicates that we are faced with one of three interesting
possibilities:

\begin{itemize}
\item The systematic or random errors in main sequence fitting
distances could be significantly larger than estimated.  This could
arise from underlying difficulties in the technique itself.
Alternately the basic data, in particular the abundances of some of
the best studied stellar systems, would have to be in serious error.

\item The predictions of theoretical stellar models could be lacking
some important ingredient, for example the change in stellar luminosity
with age or abundance.

\item The systematic errors in Hipparcos parallaxes could be
underestimated.

\end{itemize}

All three classes of solutions have important implications.  For
example, variations in the helium abundance from cluster to cluster at
fixed metal abundance could produce large changes in the distances
inferred from main sequence fitting (see for example 
Belikov et al.\ 1998).  If this is the explanation, however, large
helium abundance variations from cluster to cluster would imply an
unusual chemical evolution pattern and limit our ability to infer
distances to individual stars even when their spectral type and metal
abundance are known, possibly to the 0.3 mag level or worse.  The
problem is similar, although less severe, if the problem is in the
relative metal abundances of different systems.  If there are larger
systematic errors than claimed for the Hipparcos parallaxes, that could
affect some of the applications of the Hipparcos data.

In our view this requires a critical analysis of all methods, including
the possibility of systematic errors in the Hipparcos parallaxes.  We
will begin with the technique of main sequence fitting itself, and
follow with the Hipparcos open cluster parallaxes.

\subsection{Open Cluster Distances Based on MS Fitting}

There are two principal techniques for deriving the distances to open
clusters by main sequence fitting: relative distances to the Hyades
corrected for reddening, composition, and age; and absolute distances
relative to theoretical calculations calibrated on the Sun and nearby
stars with known parallax.  Both have their advantages and
disadvantages.

If the Hyades distance is regarded as known based upon direct distance
measurement techniques, the relative distance of other clusters can be
calculated with high precision, typically 0.02 mag or better (e.g.,
Johnson \& Knuckles 1955; Pinsonneault et al.\ 1998) if the reddening
of the other cluster is known.  One can also compute cluster distances
relative to theoretical isochrones.  Isochrones are typically
calibrated to reproduce the solar luminosity and radius at the age of
the Sun, and can therefore in some sense be regarded as anchored on the
Sun rather than the Hyades cluster.  Theoretical models predict
luminosity and effective temperature, while observers measure $V$
magnitude and color; bolometric corrections and color-temperature
relationships are therefore required to compute isochrones in the
observational plane.

Age will affect the apparent distance of a cluster.  In systems younger
than about 100 Myr the lower main sequence stars are still contracting
to the main sequence, while upper main sequence stars will be evolving
off the main sequence.  As a result, solar analogs are typically used
to measure relative distances (short pre-MS contraction times and long
main sequence lifetimes make the corrections due to cluster age small
and theoretically well-understood).  Age is therefore usually not a
major uncertainty in measuring the distance to a cluster.

Composition, however, will have a strong influence on the apparent
distance of a cluster.  In both theoretical models and direct parallax
measurements, high metallicity stars appear brighter at fixed color
than low metallicity stars.  This is partially a stellar interiors
effect; increased metal abundance makes stars of fixed mass slightly
fainter and significantly cooler, moving them above the MS locus of
lower abundance stars.  Increased line blanketing in more metal-rich
stars also makes them appear redder at fixed effective temperature.  As
a result, the apparent distance to an open cluster must be corrected
for both reddening and the metal abundance.  Changes in the helium
abundance will move models in the theoretical plane, in the sense that
a high helium abundance will make stars appear systematically fainter
at fixed effective temperature or color; however, the spectral energy
distribution and color-temperature relationship is insensitive to the
helium abundance.

The principal virtue of measuring distances relative to the Hyades is
that it relies on the slope of $M_{V}$ with respect to [Fe/H];
different empirical and theoretical calculations of color-effective
temperature relationships and bolometric corrections can disagree in
their zero-point but are typically in reasonable agreement on the
slope.  Distances relative to the Hyades are, however, subject to
zero-point shifts based on both changes in the adopted distance to the
Hyades cluster and to the metal and helium abundance of the Hyades itself.

It is important to note that the above process of main sequence fitting
is strongly based on empirical data from fundamental temperatures, the
Sun, and theoretical models which agree with helioseismic data to a
precision of order 0.1\% (Basu et al.\ 1999).

In summary, the main error sources for main sequence fitting are the
conversion from the theoretical to the observational plane, the
reddening of the cluster, and the metal and helium abundances of the
cluster.

\subsection{The Hipparcos Open Cluster Distance Scale}

The high precision and large number of stars in the Hipparcos catalog
made it possible for the first time to derive parallaxes for enough
stars in open clusters to provide trigonometric distances to systems
other than the Hyades, as well as an extremely precise distance to the
Hyades cluster itself ($m - M = 3.34 \pm 0.01$, Perryman et
al.\ 1998).  See, for example, Perryman et al.\ (1998) and van Leeuwen
(these proceedings) for a discussion of the Hipparcos
parallaxes to open clusters.

The major area of concern for the Hipparcos parallax distances to open
clusters is the possibility of correlated errors on small angular
scales.  The typical density of stars in open clusters is significantly
higher than the average across the sky; this is a known issue (e.g.,
Lindegren 1988; Narayanan \& Gould 1999b and references therein) and the
members of the Hipparcos mission attempted to correct for this
potential error source.  To explain the discrepancy between the main
sequence fitting distances and the Hipparcos distances systematic
errors of order 1 mas would be required, about ten times higher than
the Hipparcos estimates of the global systematic errors.

\section{The Problem}

For this talk we have adopted an extension of the two-color main
sequence fitting technique described in detail in Pinsonneault et
al.\ (1998).  Briefly, theoretical isochrones for stars bluer than a
$B-V$ color of 0.75 were transformed to the observational plane using
the Yale color calibration in two colors, $B-V$ and $V-I$ (Cousins).
We used the well-defined main sequence of the Pleiades to construct an
empirical isochrone for cooler stars, requiring that the distance
obtained from successive cooler bins in the Pleiades be the same as the
distance obtained for the hot stars in the Pleiades for each color.  We
applied differential corrections for the cluster age for systems with
ages different from the Pleiades.  After correcting for reddening,
distances for individual stars were computed relative to a solar
abundance isochrone and binned.  The median distance of stars within
0.2 mag of the peak in the distribution of distances was used to
determine the distance in each of two colors, $B-V$ and $V-I$.

Distances measured in $B-V$ and $V-I$ have different sensitivities to
the metal abundance; a difference between the two relative to a solar
abundance isochrones is therefore an indication of a cluster metal
abundance different from the solar value.  The difference in the $B-V$
and $V-I$ based distances was then used to estimate a photometric metal
abundance and correct the distance, adopting the sensitivity of the
color to changes in metal abundance from Pinsonneault et al.\ (1998).
Figure 1 shows the results for the open cluster NGC 2516; relative to a
solar abundance isochrone the two colors yield distinctly different
distances.  This indicates a distinctly subsolar metallicity; we derive
[Fe/H] $=-0.26$.

\begin{figure}
\plotone{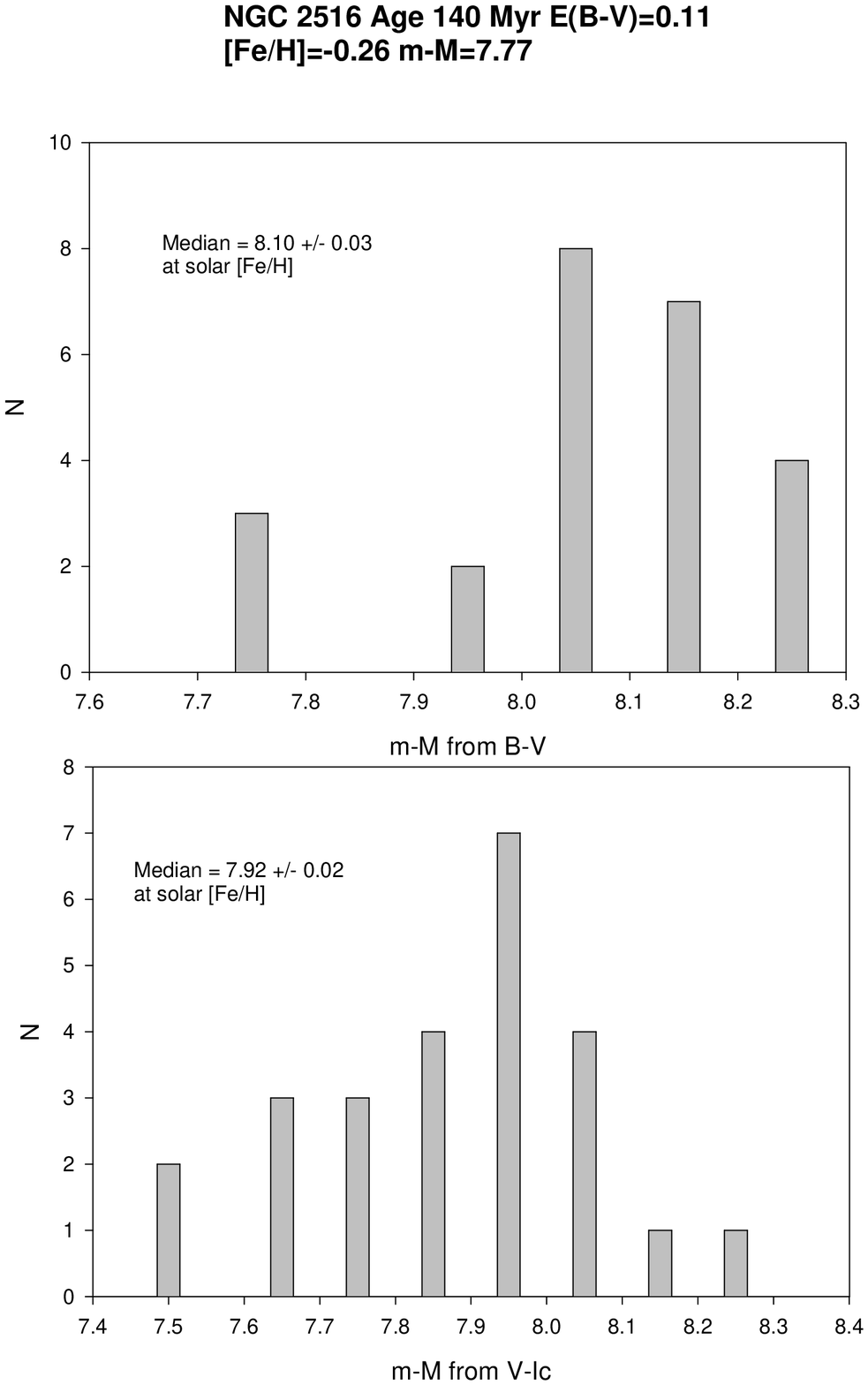}
\caption{Example of the distance and metallicity determination using
main sequence fitting in NGC~2516.  That the metallicity of this
cluster is subsolar is indicated by the derivation of a larger
distance in $B - V$ than in $V - I$ (lower metallicity brightens
the theoretical main sequence more in $B - V$).}
\end{figure}

The results from Pinsonneault et al.\ (1998) and more recent work
reported here (Pinsonneault et al.\ 2000) are compared with Hipparcos
distances from van Leeuwen (1999) and Robichon et al.\ (1999) in Table
1.

\begin{table} 
\caption{Open Cluster Distances} \label{tbl-1}
\begin{center}\scriptsize
\begin{tabular}{lrrccc}
Cluster & Age(Myr) & [Fe/H] & MSfit $m-M$ & R99 $m-M$ & vL99 $m-M$ \\ 
\tableline
${\alpha}$ Per & 60 & +0.04 & 6.28 ${\pm}0.06$ & 6.40 ${\pm}0.08$ &
6.31 ${\pm}0.08$ \\
Pleiades & 100 & +0.01 & 5.62 ${\pm}0.05$ & 5.36 ${\pm}0.07$ & 5.37
${\pm}0.07$ \\
NGC 2516 & 140 & $-$0.26 & 7.77 ${\pm}0.10$ & 7.70 ${\pm}0.16$ & $\cdots$ \\
NGC 6475 & 220 & $-$0.13 & 6.99 ${\pm}0.10$ & 7.24 ${\pm}0.19$ & 7.15 ${\pm}0.19$ \\
Coma Ber & 500 & $-$0.07 & 4.54 ${\pm}0.05$ & 4.70 ${\pm}0.04$ & 4.77 ${\pm}0.05$ \\
Hyades & 600 & +0.13 & 3.34 ${\pm}0.04$ & 3.33 ${\pm}0.01$ & $\cdots$ \\
Praesepe & 600 & +0.04 & 6.16 ${\pm}0.05$ & 6.28 ${\pm}0.13$ & 6.37 ${\pm}0.15$ \\
NGC 6633 & 700 & +0.01 & 7.94 ${\pm}0.10$ & 7.84 ${\pm}0.56$ & $\cdots$
\\ \tableline
\end{tabular}
\\
Note: R99 = Robichon et al.\ (1999).  vL99 = van Leeuwen (1999).
\end{center}
\end{table}

Note that the Hyades distance is from Perryman et al.\ (1998).  We have
adopted photometric metal abundances for ${\alpha}$ Per, the Pleiades,
NGC 2516, NGC 6475, and NGC 6633 based upon the refined main sequence
fitting technique described in Pinsonneault et al.\ (2000); for the
other clusters we have adopted the high-resolution spectroscopic
abundances of Boesgaard \& Friel (1992) and the distances computed in
Pinsonneault et al.\ (1998).  If the photometric [Fe/H] of +0.13 for
Praesepe in Pinsonneault et al.\ (1998) was adopted in preference to
the high-resolution spectroscopic abundance of +0.04 from Boesgaard \&
Friel (1992) the distance to Praesepe would rise to 6.25, close to the
shorter Hipparcos distance.  There are some systems (the Pleiades and
Coma Ber) with formally large discrepancies; there are others (NGC 6475
and ${\alpha}$ Per) which may possibly be in conflict.  There is no
obvious pattern to the deviations with age or metal abundance, and
furthermore the differences are not consistent with a simple scale
shift to systematically shorter or longer distances.  The Pleiades
discrepancy is the most striking, and as a result it has been studied
in the most detail.

\section{Error Sources for Main Sequence Fitting}

In our view none of the error sources for main sequence fitting are
large enough to account for the differences in Table 1 above.  As
discussed in previous work, neither the reddening nor systematic
differences between different sets of photometry are sufficiently
large to change the results for well-studied systems such as the
Pleiades.  We do note, however, that this may not be the case for more
distant and poorly studied systems.  Four specific areas have been
raised that could affect main sequence distances: age effects, helium
abundance variations, metal abundance differences that are larger than
inferred from main sequence fitting, and errors in the transformation
between the theoretical and observational plane.  We briefly discuss
the major objections to these explanations below.

\subsection{Age Effects}

van Leeuwen (1999) raised the possibility that there was a trend in
the discrepancy related to the age of the open clusters; this could be
an indication that the changes in stellar luminosity as a function of
age are not correctly predicted from stellar interiors models.
However, we note that differences in stellar metal abundance need to
be accounted for before comparing systems with different age.  Thus
the Hyades appears overluminous relative to other systems, but this is
predicted from main sequence fitting because it is metal-rich.  Figure
2 illustrates the difference between the present main sequence fitting
distances and those of Robichon et al.\ (1999) once the metal abundances
are taken into account; there are, for example, young systems with
main sequence fitting distances that are both closer and further away
than the Hipparcos distances.

\begin{figure}
\plotone{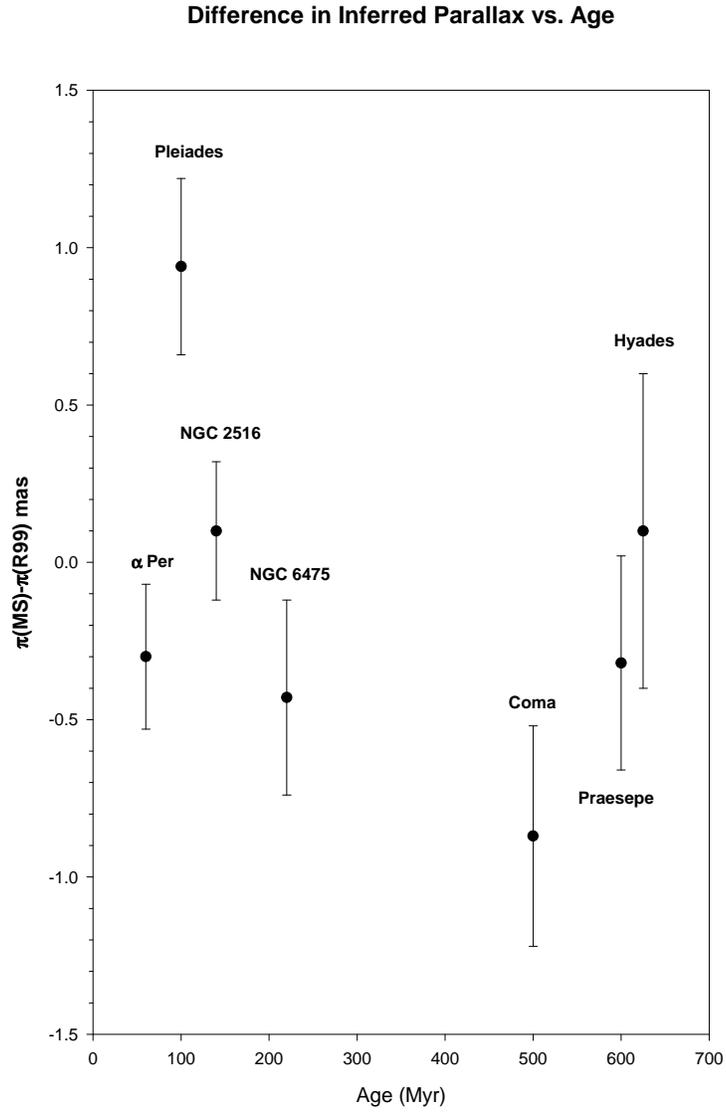}
\caption{Difference between the main sequence fitting distances
(MS) and Hipparcos distances (R99) as a function of cluster age.
The distances have been expressed as a parallax in milliarcsec.}
\end{figure}

\subsection{Helium Abundance}

Variations in helium abundance could explain the differences in Table 1
while remaining consistent with the observed metal abundances.
However, there are two principal difficulties with this solution.  The
first is that it implies an extremely unusual chemical evolution
pattern; the second is related to the relatively tight main sequence
observed for nearby field stars with high-precision parallaxes
(Soderblom et al.\ 1998).  If there was a significant population of
helium-rich solar abundance stars it would appear as a set of solar
abundance subdwarfs in the nearby field.  Soderblom et al.\ (1998)
found no evidence for solar abundance subdwarfs in a kinematically
selected sample of stars with precise Hipparcos parallaxes.  This
possibility can, and should be, tested directly in those open clusters
with members hot enough to make measurement of the surface helium
abundance possible.  However, the absence of direct evidence of high
helium abundances or of field star analogues with accurate parallaxes
makes this hypothesis an unlikely solution.

\subsection{Metal Abundance} 

A change in the open cluster metal abundances is one possible means of
explaining the discrepancy between the main sequence fitting and
Hipparcos distances.  For a number of more distant and less
well-studied open clusters high-resolution abundances are absent or few
in number and there is frequently only BV photometry available.
Praesepe may be a case where the apparent discrepancy between the
Hipparcos and main sequence fitting distance can be traced to its metal
abundance, which can be inferred from both the photometry and
high-resolution spectroscopic studies.  However, there are several
reasons why it will be difficult to reconcile the relative metal
abundances of other well-studied systems with the values implied by the
Hipparcos distances.

First, our photometric abundance scale is consistent with the
high-resolution spectroscopic abundance scale of Boesgaard \& Friel
(1992).  Relative abundances can be derived more accurately than
absolute ones, and it is the relative heavy element abundance and not
the absolute abundance that is important for determining relative
distances.  Furthermore, in some cases (such as the Pleiades) the
required differences are large.

We have demonstrated that even modest departures from solar abundance
can easily be measured by comparing distances inferred from metallicity
sensitive and metallicity insensitive colors.  We see no evidence for a
subsolar Pleiades metal abundance.

Additional colors with different metallicity sensitivities would be
useful to test the metallicity scale.  For example, Robichon (these
proceedings) inferred a larger difference in the metal abundances of
the Pleiades and Praesepe from Geneva photometry than from $BVI$
photometry.  In this case, however, the distances derived from other
color indices were not consistent for the adopted relative metal
abundances.  The inclusion of other color indices with well-calibrated
metallicity sensitivities could be an important test of the robustness
of photometric metallicities.  In summary, changes in the metal
abundances of clusters would significantly alter the distances, but
there are several lines of evidence that argue against such a solution
in the most problematic cases.

\subsection{Color Transformations}

An additional possibility is raised by Robichon (these proceedings):
the absolute color calibrations may themselves be an error source.  In
addition, Robichon et al.\ (1999) pointed out that different
investigators have obtained very different distances to the same open
clusters.  This indicates that at least in some cases the systematic
errors in main sequence fitting have been underestimated.

Changes in the color-temperature relationship and bolometric
corrections would manifest themselves as a shift in the zero-point of
the cluster distance scale.  This can be tested by comparing the
isochrones with the Hyades distance; our method produces a Hyades
distance in agreement with the Hipparcos distance for a solar-scaled
helium abundance.  Other transformations from the theoretical to the
observational plane (e.g., Perryman et al.\ 1998) require a lower helium
abundance for the Hyades to reproduce the distance inferred from other
methods.  In our view this ends up substituting one problem (the
Pleiades distance) for another (explaining the low inferred Hyades
helium abundance.)  In addition a systematic shift to lower distances
would create tension between the Hipparcos and main sequence fitting
distances to other clusters.

We agree with Robichon (these proceedings) that in some cases the
systematic errors from main sequence fitting have been underestimated
in previous studies.  Different Hyades zero-point distances, neglecting
cluster-to-cluster variations in metal abundance, deriving distances in
only one color, and usage of the upper main sequence rather than the
lower main sequence for distance estimates can all generate
study-to-study scatter or even in some cases shifts in relative cluster
distances.  However, the precision of each of the individual studies
must be examined to reliably infer their individual errors.

\section {Possible Systematic Errors in the Hipparcos Parallaxes}

Pinsonneault et al.\ (1998) analyzed the Hipparcos parallax data and
found some evidence that suggested the systematic errors could be 
underestimated on small angular scales.  Narayanan \& Gould (1999ab)
used a radial velocity gradient technique to infer the distances to
the Hyades and Pleiades; in that study the deviations between the
Hipparcos parallaxes and the stellar distances inferred from their
method were found to exhibit spatial correlations.  The idea that
there are systematic errors in the Hipparcos parallaxes that cover
large areas of the sky has been energetically challenged (see for
instance Robichon et al.\ 1999.)  More work needs to be done,
especially involving other distance estimation techniques that can
serve as independent tests of both main sequence fitting and
systematic errors in the Hipparcos parallaxes at small angular scales.

\section {Things to be Done}

Quite aside from any connection to parallaxes, it is of value to extend
the technique of main sequence fitting to more (and more distant) open
clusters.  The biggest need here is to obtain high quality multicolor
photometry, from which the cluster metallicity and age (in certain age
ranges) can be determined.   Many potentially interesting clusters
still have rather large distance errors because of incomplete
membership or a lack of good estimates of the foreground extinction.

Finding better ages and distances bears upon many areas of research
besides the obvious ones of calibrating standard candles (e.g.,
Cepheids) and of the evolution of the open cluster system.  For
example,  some of the conclusions from open cluster studies such as
lithium depletion depend on age and metallicity, not known well for
many clusters.

Finally, we further that any new independent tests of errors in the
Hipparcos parallaxes be performed.

We gratefully acknowledge the efforts of Jeremy King, John Stauffer,
and Robert Hanson in this investigation.

\end{document}